\documentclass{aa}  
\usepackage{graphicx}
\usepackage{txfonts}
\usepackage{natbib}
\bibpunct{(}{)}{;}{a}{}{,} 
\newcommand\obsline{C~{\sc iii}~97.7~nm}

\begin{document}
   \title{Search for photospheric footpoints
of quiet Sun transition region loops}


   \author{J. S\'anchez Almeida\inst{1}\and
	L. Teriaca\inst{2}\and
	P. S\"utterlin\inst{3} \and
	D. Spadaro\inst{4}\and 
	U. Sch\"uhle\inst{2}\and 
	R. J. Rutten\inst{3}
	}
  
   \offprints{jos@iac.es}

   \institute{
	Instituto de Astrof\'\i sica de Canarias, E-38205 La Laguna, Tenerife, Spain\\
	\email{jos@iac.es}
	\and
	Max-Planck-Institut f\"ur Sonnensystemforschung, 
	Max-Plank Str. 2, 37191 Katlenburg-Lindau, Germany\\
	\email{teriaca@mps.mpg.de, schuehle@mps.mpg.de}
	\and
	Sterrenkundig Instituut, Universiteit Utrecht, P.O. Box 80000, 
	NL-3508 TA Utrecht, The Netherlands\\
	\email{P.Suetterlin@astro.uu.nl, R.J.Rutten@astro.uu.nl}
	\and
	INAF-Osservatorio Astrofisico di Catania, I-95123 Catania, Italy\\
	\email{dspadaro@oact.inaf.it}
            }

   \date{Received 20 June 2007 ; accepted 18 September 2007}

 
  \abstract
   {
	The footpoints of quiet Sun Transition Region (TR) loops do not 
	seem to coincide with the photospheric magnetic structures 
	appearing in  traditional 
	low-sensitivity 
	magnetograms.
	}
   {To look for the so-far unidentified photospheric footpoints of TR loops using G-band bright points
	(BPs) as proxies for photospheric magnetic field 
	concentrations.
	}
   {Comparison of TR measurements 
	with SoHO/SUMER and photospheric magnetic field observations 
	obtained with the Dutch Open Telescope.}
   {Photospheric BPs are associated with bright TR structures, but
	they seem to avoid the brightest parts of the structure. 
	BPs appear in regions that are globally redshifted, but they avoid extreme velocities.
	TR explosive events are not clearly associated with BPs.
}
   {

The observations are 
not inconsistent with the BPs being footpoints of TR loops,
although
we have not succeeded to uniquely identify particular
BPs with specific TR loops. 
}


\authorrunning{S\'anchez Almeida et al.}
\titlerunning{Photospheric footpoints
of quiet Sun TR loops}

   \maketitle
%

\section{Rationale}\label{introduction}

The solar Transition Region (TR) is defined as the part of the solar
atmosphere characterized by temperatures from $2\times10^4$~K (upper
chromosphere) to 10$^6$~K (corona), and densities 
from $\sim10^{10}$~cm$^{-3}$ to $\sim10^8$~cm$^{-3}$
\citep[e.g.,][]{mariska92}.
In classical 1-D coronal models 
\citep[e.g.,][]{gab76}, the TR is a thin ($\sim\,$100 km) thermal interface
between the cooler chromosphere and the hotter corona. 
Although such {\em interface} regions must exist at the footpoints of 
large active region coronal loops,
they appear to be responsible of only a small fraction of the TR
emission  
\citep[e.g.,][]{ath82,fel83}.
The quiet Sun TR, in particular, does not represent a 
continuous transition between the chromosphere and
the corona. Rather, sensitive UV observations 
show the upper solar atmosphere to consist of 
a hierarchy of loop structures with different temperatures
and extents \citep[e.g.,][]{dow86,fel00,fel02}. 
Small cool looplike structures fill most of the quiet Sun 
images and spectroheliograms obtained in lines formed at TR 
temperatures  \citep[see][Fig.~7]{fel99}.
The footpoints of such loops 
do not seem to be 
associated with known traditional
magnetic structures. 
They lie across network boundaries
with the footpoints presumably 
in the interior of supergranulation cells. 
\citet{fel01}  find no chromospheric counterpart 
near the apparent footpoints of the 
structures.
\citet{war00} find that the loops do not connect 
magnetic structures in 
full-disk magnetograms obtained with the Michelson
Doppler Imager (MDI) aboard SoHO.
This kind of magnetogram, however, does not have enough 
spatial resolution and sensitivity to reveal magnetic
structures in supergranulation cell interiors. In fact, it has
been known for a long time that such structures do exist 
\citep{liv75,smi75}.
They show up as weak Hanle depolarization
signals \citep[e.g.,][]{ste82,fau93,tru04},
weak Zeeman polarization signals 
\citep[e.g.,][]{wan95,lin99,san00,dom03a}, and small bright 
points in intergranular lanes \citep{san04a,dew05}.
According to numerical simulations 
\citep[e.g.,][]{cat99a,vog05,vog07,stei06} and
observations \citep[e.g.,][]{san00,san03,dom06},
a complex magnetic field pervades 
the seemingly non-magnetic {\em quiet} photosphere.
Theoretical arguments suggest that
a significant part of such photospheric magnetic
field actually reaches 
the TR and the corona
\citep{sch03b,jen06}. 
Obviously, these ubiquitous 
magnetic fields seem to be 
the natural candidates for the so-far unidentified
quiet Sun TR loop footpoints.  If such conjecture turns out to be correct,
it offers a new standpoint for studying and 
understanding the nature of the TR and its loops. In addition,
it would provide
a new scientific rationale for studying
the magnetism of the quiet Sun.
Guided by these ideas, we undertook a first exploratory
study to identify the photospheric footpoints of the 
quiet Sun TR loops. Such work is described in the present paper.

\begin{figure}[!t]
\includegraphics[width=8cm]{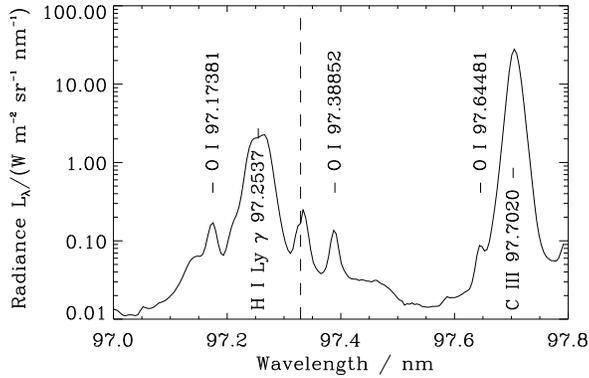}
\caption{
Spectral range of the SUMER observations 
averaged over the full FOV. Spectral radiances
are given in W~m$^{-2}$~s$^{-1}$~sr$^{-1}$~nm$^{-1}$.
All relevant lines are indicated with the known rest wavelengths 
(O~{\sc i} from \citeauthor{kelly87} \citeyear{kelly87}; 
C~{\sc iii} and H~{\sc i}~Ly~$\gamma$ from 
\citeauthor{mort03} \citeyear{mort03}). 
The vertical dashed line marks 
the boundary between the KBr-coated and the bare sections of the SUMER 
detector.}
\label{spectrum}
\end{figure}

	The study requires simultaneous 
observations of the quiet Sun TR and the photospheric
magnetic fields. The Solar Ultraviolet Measurements of Emitted Radiation 
(SUMER) spectrometer \citep{wil95} aboard SoHO was used to record the 
97~nm to 98~nm spectral range 
(Fig.~\ref{spectrum})
that includes the H~{\sc i}~Ly~$\gamma$ 97.2~nm line 
($T\sim1.5\times10^{4}$~K) and the \obsline\ line ($T\sim8\times10^{4}$~K).
The latter is one of the brightest lines of the solar VUV spectrum, allowing
low-noise spectra to be recorded with exposure times of a few seconds.
Being formed in the middle TR, the \obsline\ line is an ideal tracer
of the TR 
looplike structures
\citep[][]{fel99}.
As proxy for the quiet Sun
magnetic fields, we employ high spatial resolution images
in the so-called G~band (the CH molecular band at 
430~nm). G-band bright points (BPs)  
in intergranular
lanes are proxies for intense kG magnetic concentrations
\citep{mul84,ber95,ber98b}. The reasons for using G-band BPs
as magnetic tracers is twofold. Magnetic structures having
kG fields represent only a very small fraction of the quiet
Sun magnetic structures, which have field strengths in
the full range from 0~kG to 2~kG \citep{san00,soc02}. However,
the magnetic loops rooted in kG concentrations are 
expected to tip-over higher-up \citep{dom06},
and so they have the largest probability of 
being the footpoints we are seeking. 
The second reason has to do with feasibility --
unpolarized imaging is simpler than spectro-polarimetry,
and G-band BPs are easy to detect provided that the
imaging has enough spatial resolution \citep{tit96}.
We employ the 45~cm Dutch Open Telescope (DOT) which, using
speckle image reconstruction techniques, can provide
time series of
diffraction limited images in a routine fashion
\citep[][]{rut04b}. 
%

The paper is organized as follows. The observations,
reduction, and alignment are described in \S~\ref{observations}.
Aligning the ground-based images with the satellite images is 
critical, which justifies the details given in \S~\ref{observations}.
The observational results are put forward in 
\S~\ref{obs_res}. The implications of such results are 
analyzed in \S~\ref{conclusions}. 
	
\section{Observations, data reduction and co-alignment}\label{observations}

The observations were carried out on March 25, 2006, from
9:00~UTC to 11:00~UTC. Due to poor weather conditions 
at the DOT site, 
it turned out to be the only useful time slot out of
three attempts in a coordinated campaign involving SUMER,
MDI \citep{sch95b}, 
and DOT.
This section describes the data sets plus the
procedure that allowed us to bring the satellite 
images and the ground-based images to a common reference
system.

\subsection{SUMER spectra and MDI magnetograms}\label{sumerobs}

\begin{figure*}
\centering
\includegraphics[angle=90.,width=0.9\textwidth]{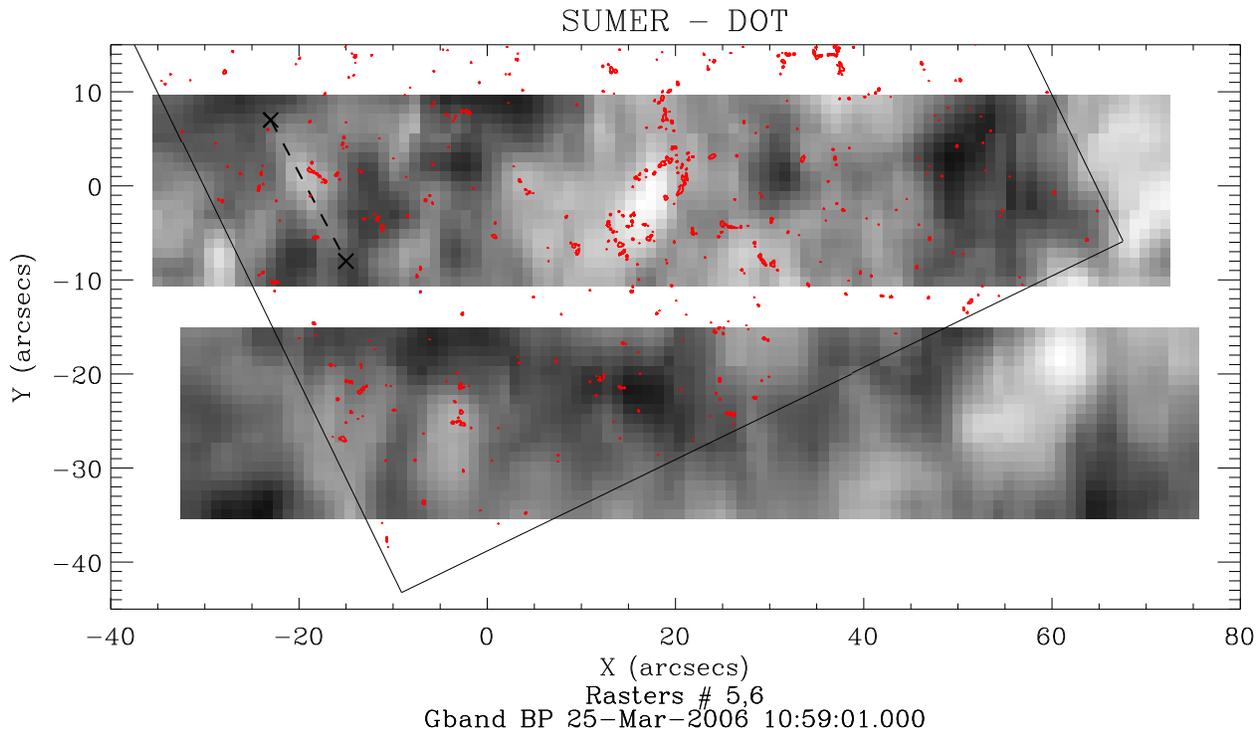}
\caption{
SUMER \obsline\ radiances of rasters 5 and 6 (the image) together with
the position of the G-band BPs (the red contours). 
Coordinates are referred to the solar disk center 
as provided by the MDI magnetograms. 
The \obsline\ radiances are shown in a logarithmic scale.
The rotated box outlines the DOT FOV.
The dashed line limited by times symbols illustrates
the kind of elongated structure that may be  ascribed to
individual loops projected on the solar surface.
}
\label{res3}
\end{figure*}

The SUMER instrument is a slit spectrometer and, therefore, images
are obtained by raster scanning across the region of interest.
Our data consist of six rasters taken sequentially and forming three pairs.
Due to a failure of the SUMER 
A detector,
only a strip of about 20\arcsec\ along 
the slit could be imaged and, therefore, a second raster scan was placed 
25\arcsec\ towards the south of the previous raster. The same region near the 
center of the solar disk was observed three times -- always in east-west 
direction -- yielding the six scans mentioned above. Each single raster is made 
of 99 step positions taken with a cadence of 13~s (12~s exposure) and 
therefore lasting about 22 minutes.
Between rasters the solar rotation was compensated automatically by
displacement of the field-of-view (FOV) towards west. One pair of rasters
renders a FOV of about 
100\arcsec~$\times$~45\arcsec (see Fig.~\ref{res3}).
After accounting for solar rotation, the step size 
of the scan turns out to be 1\farcs 092, which is similar
to the sampling interval in the direction along the slit  
(1\farcs015).
SUMER has a spatial resolution of 1.5\arcsec\ \citep[][]{lem97}, corresponding
to about 1000 km on the Sun at the distance from SoHO to the Sun.

%
%
The 
spectra were corrected for flat-field inhomogeneities and for the 
geometrical distortion induced by the electronic read-out scheme by using the
standard routines available to SUMER users. 
We corrected for residual geometrical 
distortion in the spectral direction, that is significant at the edges of
the detector. This was done by {\em straightening} 
the averaged (over the entire dataset) slit spectrum around the O~{\sc i}
lines present in the observed spectral range (see Fig.~\ref{spectrum}). 
Lines from neutrals and single-ionized species are 
known 
to show very 
small average Doppler shifts and can be used to obtain a wavelength
calibration \citep[e.g.,][and references therein]{ter99}. 
Finally, the
data were 
calibrated to yield spectral radiances in W~m$^{-2}$~sr$^{-1}$~nm$^{-1}$.
The recorded spectral range
is shown in Fig.~\ref{spectrum}.
The spectra 
were  
used to compute integrated radiances 
at selected bandpasses together with Doppler velocities, and 
line widths. 
Three bandpasses are mentioned in the paper: 
\obsline\ (97.65~nm -- 97.75~nm), 
pseudo-continuum (97.4~nm -- 97.6~nm),
and 
H~{\sc i}~Ly\,$\gamma$
(97.2~nm -- 97.3~nm).
In order to obtain velocities 
and line widths, the line profiles of \obsline\
and the underlying continuum
were fitted with a single Gaussian plus
a second-order polynomial. 
The procedure provides errors for the velocities and
Doppler widths. The mean errors
considering the full FOV turn out to be 1.5~km~s$^{-1}$
and 0.5~pm, respectively.  
When a rest wavelength of 97.702~nm is used \citep{mort03},
the average shift over the entire dataset is 11~km~s$^{-1}$. 
This value is
comparable to the values found in the 
literature for lines formed at similar temperatures 
\citep[see e.g.,][and references therein]{ter99}.
The structures appearing in the SUMER \obsline\  maps  
have dimensions very similar to those described by
\citet{fel99,fel00} 
as cool loop-like structures clustering across the 
chromospheric network boundaries. The 
structures are less evident, 
but this is mostly due to our much smaller FOV,
as compared to the images published by Feldman and 
coworkers (270\arcsec$\,\times\,$300\arcsec).
This fact, together with the lack
of an objective definition of \obsline\ loop-like 
structure, makes it uncertain identifying  
individual loops in our maps. One can find features in Fig.~\ref{res3} 
that may be  ascribed to individual loops projected on the solar 
surface (see the example shown as a dashed line
limited by two times symbols), but such identification is 
not free from ambiguity. 
Because of this reason, we will use the term 
{\em elongated bright structure}
to describe the elongated patchy structures in Fig.~\ref{res3},
keeping in mind, however, that they should be identified with
the structures termed loops by \citet{fel99}.


In addition to the SUMER data, we also use full disk SoHO/MDI 
magnetograms taken from 9:00 UTC to 11:30 UTC with a cadence of 1~min. 
In this case the pixel is 1\farcs 98 square, with the noise level 
corresponding to 
some
16~G \citep{sch95b,liu01}.

\subsection{DOT images} 

\begin{figure*}
\centering
\includegraphics[angle=90.,width=\textwidth]{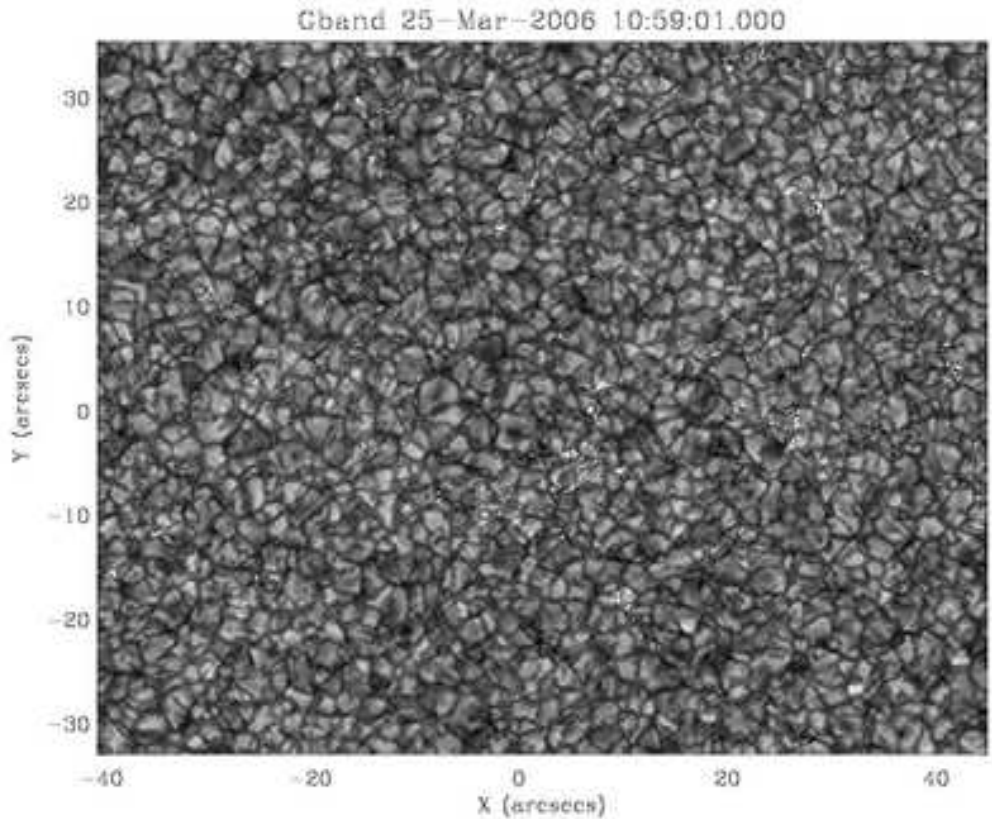}
\caption{
Example of speckle reconstructed G-band image.
It shows many G-band bright points tracing magnetic concentrations
in intergranular lanes. 
The spatial coordinates are referred to the center 
of the image. 
}
\label{gband1}
\end{figure*}

This work employs only part of the series of images routinely
provided by the DOT  \citep{rut04b}, in particular,
we analyze G-band images, 
and Ca {\sc ii} H line core images. They are restored 
using speckle techniques \citep{rut04b}   which, under good seeing conditions,
render diffraction limited images 
($\sim$0\farcs 2 
in the G~band). 
The final images have a sampling interval of  0\farcs071, with a FOV 
of 85\arcsec~$\times$~70\arcsec~ (see Fig.~\ref{gband1}).
The time series has an irregular cadence equivalent 
to one image every two minutes. The best seeing  
occurs by the end of 
the time sequence and, therefore, it is associated with 
SUMER rasters 5 and 6 (Fried's parameter 
8.5 cm). 

As expected,  the G-band images are
full of BPs in intergranular lanes (Fig.~\ref{gband1}).
Automatic identification of BPs is not trivial since
a simple threshold criterion does not suffice.
Sometimes granules are brighter than BPs clearly visible
in the intergranular lanes \citep[e.g.,][]{bov03}. We applied a simple
algorithm consisting of three steps:
(1) construction of a smoothed version of the G-band image
which is computed after removal of the brightest features 
in the image, (2) subtraction of the smoothed image from the
original one to enhance the small bright features,
and (3) selection of the bright features in the subtracted
image, but only when they are localized in dark areas 
of the smoothed image.
The first step, where the brightest features are removed, produces
a smoothed version of the image which is not contaminated by the
presence of BPs. The difference between the full image and 
this smoothed version enhances the contrast of the G-band BPs, 
and this high-contrast image is used to select the bright features
existing in the dark intergranular lanes.
The algorithm does a good job,
in the sense that it agrees with the visual identification
of the BPs. It is not perfect, and a
few bright borders of granules are misidentified as
BP, and some BPs are overlooked. However,
the identification suffices for the exploratory 
analysis carried out in this paper.
The use of a different algorithm would slightly
modify the number of selected BPs. 
As we explain in \S~\ref{obs_res},
the trends that we obtain remain
the same for all the three SUMER raster pairs, which 
correspond to different seeing conditions at the 
DOT site.  Since seeing modifies the number of detectable 
BPs, and it does not change the trends,
the details on how the BPs are detected 
do not seem to alter our results.

\subsection{Alignment}
The quiet Sun magnetic network is well defined in
both standard magnetograms and Ca~{\sc ii} line core filtergrams 
\citep[e.g.,][]{bec77b}. Our method  uses
this property to align satellite data (SoHO/MDI magnetograms) with 
ground-based data (DOT Ca~{\sc ii}~H images). 
We bring both SUMER raster scans and
DOT G-band images to the spatial coordinates of 
MDI. Then SUMER raster scans and DOT G-band 
images can be superposed directly. The actual procedure
is explained  
here in some detail.

\begin{figure*}
\includegraphics[angle=90,width=0.9\textwidth]{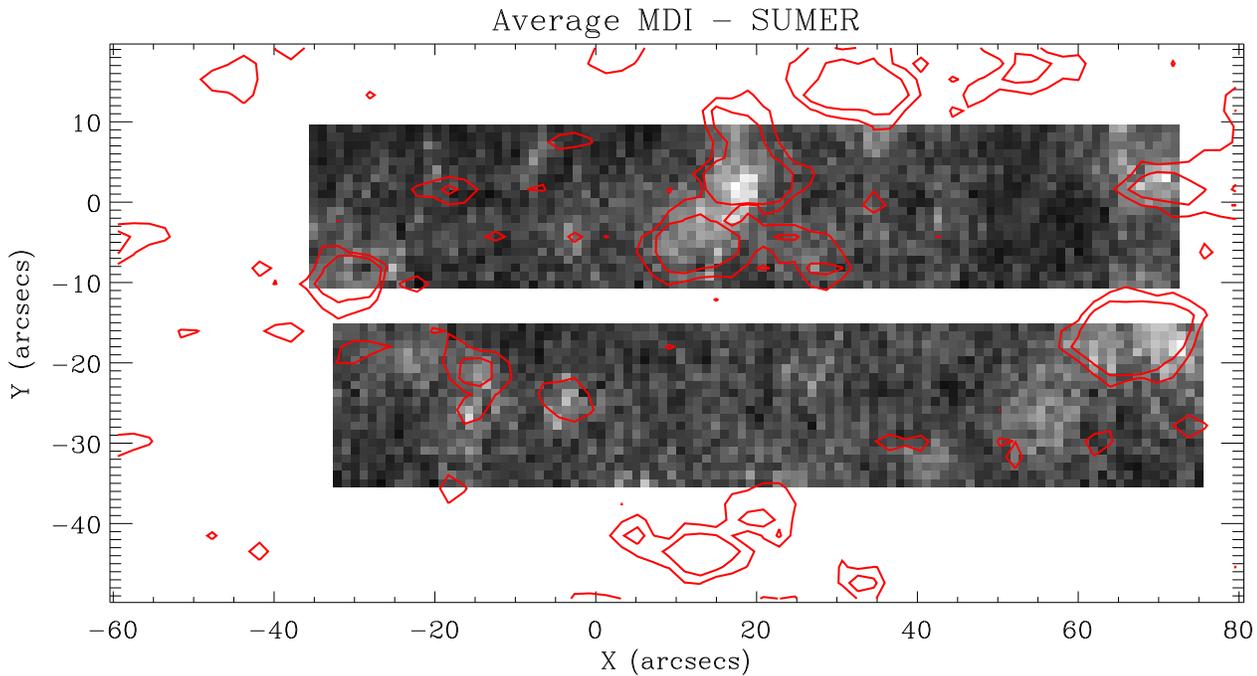}
\caption{
Example of alignment between a SUMER 
pseudo-continuum image and the absolute
value of the average MDI magnetogram. 
The latter is shown as a contour plot, with contours at 7~G and 15~G.
The spatial coordinates are referred to the solar
disk center according to the MDI scale.
}
\label{sumer_mdi}
\end{figure*}

Although SUMER data and MDI data come from a 
single
satellite, they are not co-aligned. Errors
in the raster mechanism and thermal excursions
of the payload produce an unpredictable offset
of up to 10\arcsec. The SUMER to MDI co-alignment
has been accomplished by comparing the pseudo-continuum
SUMER image coming from the bandpass between 97.4~nm and 
97.6~nm (Fig.~\ref{spectrum}), with the average among
the MDI magnetograms taken during the time-span of the
SUMER raster scan. The co-alignment is carried by trial
and error,  
blinking 
on a computer screen 
the images of the SUMER rasters and
the absolute value of the average MDI magnetogram.
One of the images is then
shifted with respect to the 
other to get the best match. 
The results are illustrated in Fig.~\ref{sumer_mdi}.
Repetitions of the exercise always yield
the same offset within $\pm$ 1\arcsec (optimistic view),
and $\pm$ 2\arcsec (conservative view). 
If the procedure uses Ly\,$\gamma$ photons
instead of
pseudo-continuum, then the alignment
remains the same within the quoted 
uncertainties. 
We also tried aligning with and without removal of the solar rotation
when averaging the magnetograms, and using a logarithm 
grayscale
to represent SUMER radiances. No significant change is observed.
The three couples of SUMER raster scans give 
the same offset.

The alignment between DOT and MDI is carried out
by means of the Ca {\sc ii} H images.
DOT images are both rotated and shifted
with respect to SoHO images. The rotation is given
by the angle between the geocentric 
North and the solar rotational North, 
which we set according to the ephemeris.
As for the relative displacement, we use the same trial
and error approach described above for the  
SUMER~to~MDI alignment. Since 
the resolution of the MDI magnetograms is much lower
than the speckle restored DOT Ca {\sc ii} H images,
the average among the burst of images 
gathered for speckle restoration is used for comparison.
Errors are smaller than the  
SUMER~to~MDI alignment since the structures 
observed in DOT Ca {\sc ii} H and MDI are quite similar
and 
therefore 
easy to identify.
After repeating the 
trial and error process several times, one finds the 
displacements to be consistent within 1\arcsec.
DOT G-band images and  DOT Ca {\sc ii} H are also 
misaligned. We find the shift between the images by
cross-correlation. This method does not correct for the 
slight different 
orientation of the two images, and for a 
small difference of the spatial scales. 
However, the two effects leave a residual error 
always well below 1\arcsec.

In short, the critical part of the alignment
has been carried out by trial and error and, therefore, its
uncertainty is difficult to estimate. However,
judging the errors by the consistency 
of the trial and error process, they should
be smaller than 2\arcsec.
This uncertainty is mostly set by the 
SUMER~to~MDI alignment, making all other 
errors negligible.


\section{Observational results}\label{obs_res}

The best DOT seeing occurred by the end of the time series,
corresponding to SUMER rasters 5 and 6.
The number of G-band BPs in an image 
depends critically on the spatial resolution 
\citep{tit96,san04a}, therefore, our analysis is
focused on these last rasters and the best G-band image 
taken together with them. We also
analyzed the other raster pairs, and other snapshots
of the time series. The results are always consistent
with those reported below.

\subsection{Radiances}\label{results1}

\begin{figure*}
\centering
\includegraphics[angle=90.,width=\textwidth]{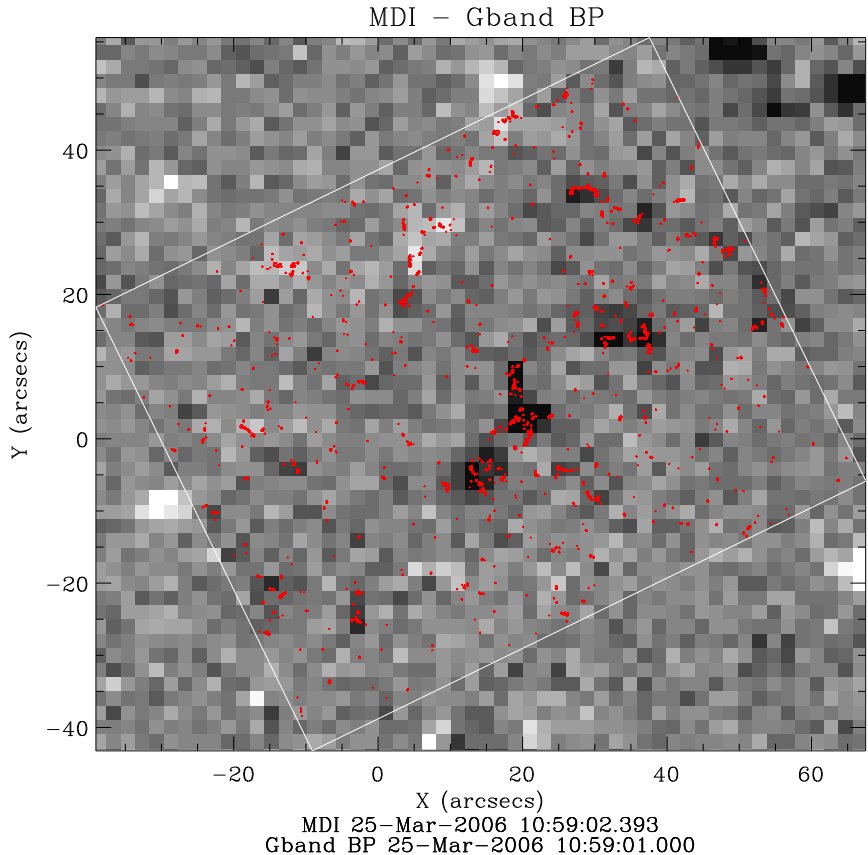}
\caption{
MDI magnetogram (the image) together with 
the BPs existing in the G-band image (red contours).
The magnetogram has been scaled from -20~G to 20~G, so that
only the black and white pixels are well above the noise
level ($\sim$16~G). 
The coordinates are referred to the MDI solar disk 
center. The rectangle indicates the DOT G-band FOV.
Note that these two images have not been aligned directly 
but via Ca~{\sc ii}~H images
}
\label{gband4}
\end{figure*}

\begin{figure}
\centering
\includegraphics[angle=90,width=9cm]{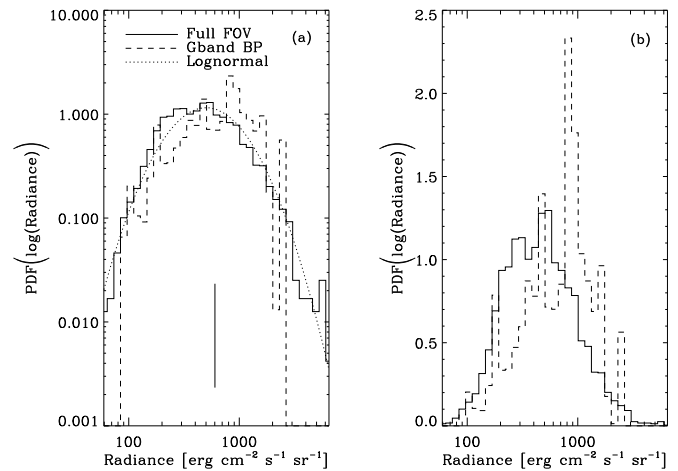}
\caption{
(a) Histogram (PDF) of the logarithm of the SUMER \obsline\ 
radiance corresponding to the full SUMER FOV
(the solid line), and 
at the pixels having G-band BPs (the dashed line).
The dotted line corresponds to a
lognormal approximation of the full FOV PDF.
The vertical solid line indicates the mean radiance of the full
FOV. (b) Same as (a) but showing the PDFs in a linear scale
to emphasize the systematic increase of radiance in those
pixels with BPs.}
\label{histogram1}
\end{figure}

The G-band images show a significant number of BPs 
with no obvious counterpart in the MDI magnetograms 
(Fig.~\ref{gband4}). As we point
out in \S~\ref{introduction}, a significant number of 
photospheric magnetic structures, and so of TR loop
footpoint candidates, 
does 
not show up in traditional 
measurements.
The G-band BPs existing in the
DOT image of best angular resolution (Fig.~\ref{gband1})
are overplotted on the SUMER \obsline\ image  in Fig.~\ref{res3}. 
Visual inspection indicates that the photospheric
BPs are associated with bright  \obsline\ features, although 
they seem to avoid the brightest cores.
This is particularly clear for the case of the brightest 
\obsline\ structure in  the SUMER FOV, located 
at $X\simeq 15$\arcsec\ and $Y\simeq 0$\arcsec\
(see Fig.~\ref{res3}).
In order to quantify this impression, we computed the 
histogram of \obsline\ radiances for the full SUMER FOV, and
for the \obsline\ radiances at the 
positions of the G-band BPs (i.e., assigning the radiance of the closest 
SUMER pixel to each DOT pixel classified as G-band BP).
The result is shown in Fig.~\ref{histogram1}, where
the histograms have been normalized 
to one
and termed PDF 
(acronym for probability density function). 
As expected \citep[e.g.][]{pau00}, the PDF of the full FOV
is approximately lognormal (a parabola in a log-log plot; the dotted line 
in Fig.~\ref{histogram1}a).
%
%
The PDF of \obsline\ radiances 
at the G-band BPs turns out to be shifted toward large 
radiances (see the dashed line). This shift is a feature 
common to all three pairs of raster scans. Note also the lack of 
very bright features in the G-band histogram --  
the dashed line in  Fig.~\ref{histogram1}a drops
down for radiances larger than 2500~mW~m$^{-2}$~sr$^{-1}$.  
This drop may unveil 
the  tendency for the BPs to avoid the core of the bright 
\obsline\ structures.

\subsection{Velocities and Doppler widths}\label{results2}

\begin{figure*}
\includegraphics[angle=90.,width=0.9\textwidth]{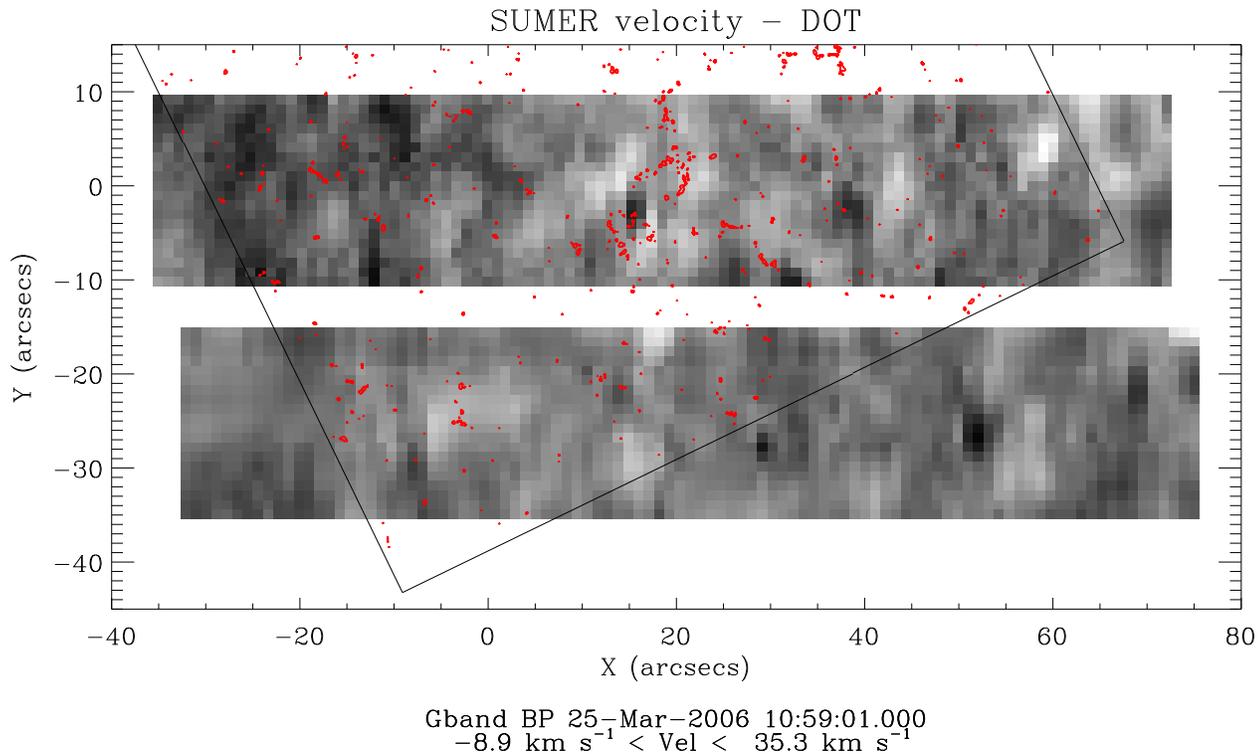}
\caption{
\obsline\ velocities (the image) 
together with the position of the G-band BPs (the red contours). 
The spatial coordinates and the box are the same as for Fig.~\ref{res3}. 
The image has been scaled between the minimum 
and maximum values indicated in the subtitle, with the white 
color representing the largest redshift.
}
\label{vel56}
\end{figure*}

\begin{figure}
\includegraphics[angle=90.,width=9cm]{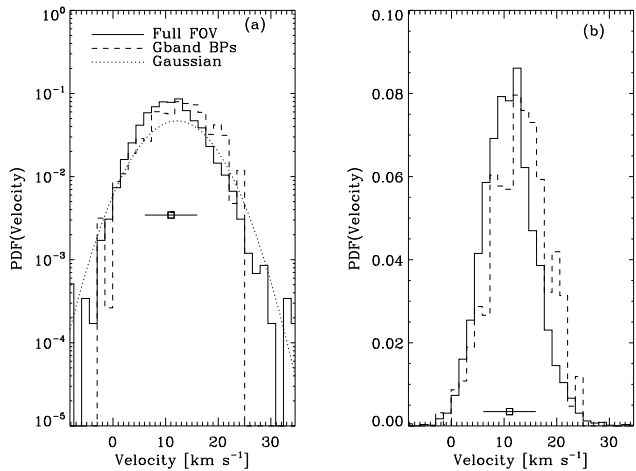}
\caption{
(a) Histogram (PDF) of the distribution of \obsline\ velocities
shown in a logarithm scale.
As indicated by the inset, the solid line corresponds
to the full SUMER FOV, whereas the dashed line shows the 
distribution of velocities in those places with G-band BPs.
The dotted line corresponds to a Gaussian fit to the full FOV
PDF.
(b) Same histograms as in (a) but shown in a linear scale 
to  emphasize the global redshift of the SUMER spectra
in pixels with G-band BPs.
Positive velocities correspond to redshifts.
The square symbols with horizontal error bars
represent the average plus-minus the standard
deviation of the full FOV velocities.
}
\label{vel_his}
\end{figure}

Figure~\ref{vel56} contains the map of velocities derived from \obsline\ 
together with the G-band BPs.
%
The bright and dark patches of the
velocity image are devoid of BPs,
which avoid extreme velocities.
This view is corroborated by the 
histogram of velocities presented in Fig.~\ref{vel_his}.
It shows the histograms (PDFs) for the full 
SUMER FOV (the solid lines), together with the 
histograms of velocities at the position of the 
G-band BPs (the dashed lines). The BP histogram
lacks of the largest blueshifts and 
redshifts existing 
in the full FOV histogram.
%
Another property easy to extract from Fig.~\ref{vel_his}b
is the global redshift of the SUMER spectra with
BPs with respect to the full FOV velocities. 
The mean and standard deviation of the velocity 
distributions are
$11.0\pm 5.0$~km~s$^{-1}$ (full FOV) and  
$12.9\pm 5.3$~km~s$^{-1}$ (BPs), which 
renders a relative
shift of the order of 1.9\,km~s$^{-1}$.
The existence of such excess of redshift
is a very robust result.
The velocity histograms are made of hundreds of points, 
and they are well represented by Gaussians with widths of about 
5 km~s$^{-1}$.
The histograms of the means are necessarily much narrower,
with widths of the order of 5 km~s$^{-1}$ divided by 
the square root of the number of individual velocities used 
to compute the means \citep[e.g.,][]{mar71}. In our case this 
uncertainty of the means ($\le$0.3 km~s$^{-1}$) is much smaller 
than the separation between the means.
Two final comments are in order. First, the
difference of velocities cannot be caused 
by systematic errors biasing our wavelength calibration,
since they would affect the two histograms in the same
way. Second, similar
shifts are present in all raster scans,
reinforcing the results.

\begin{figure*}
\centering
\includegraphics[angle=90.,width=0.9\textwidth]{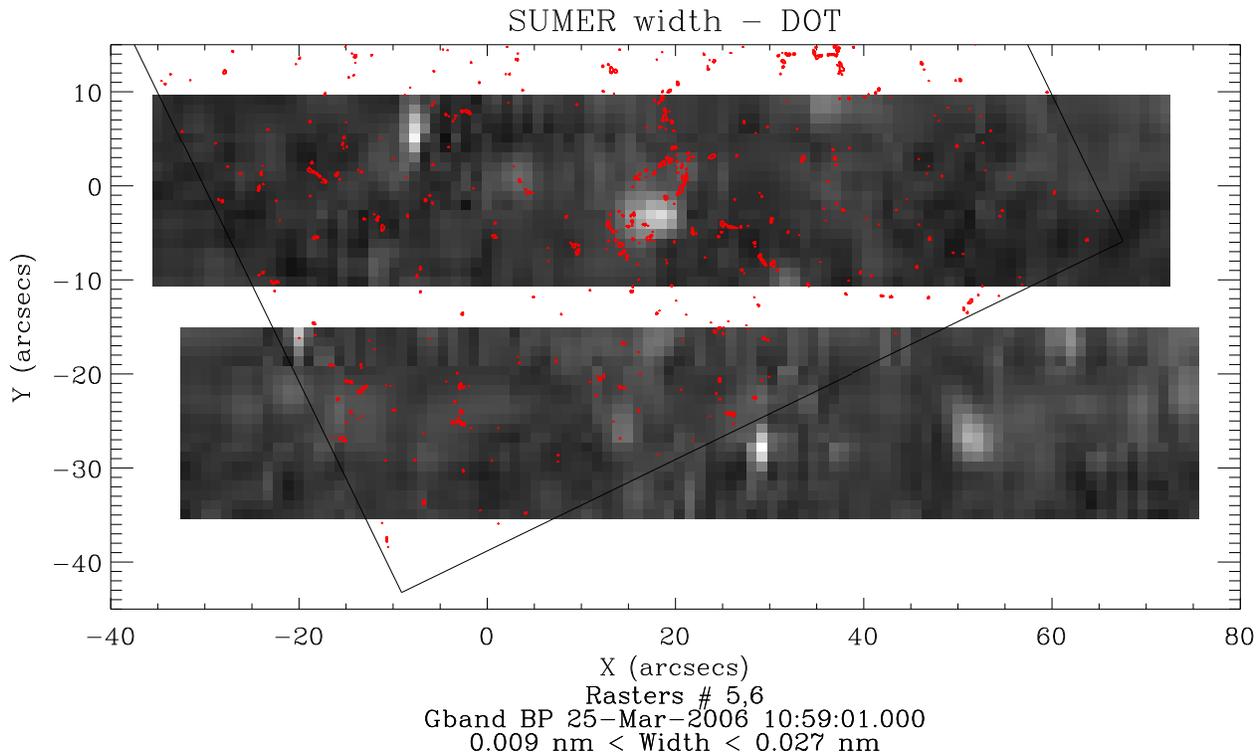}
\caption{
\obsline\ line widths in raster scans 5 and 6 (the image) 
together with the position of G-band BPs (the red contours). 
The spatial coordinates and the box are the same as for Fig.~\ref{res3}.
The image has been scaled between the minimum 
and maximum widths indicated in the subtitle.}
\label{widths}
\end{figure*}

Nothing special seems to be associated with the maps of \obsline\ widths
except, perhaps, the fact that the largest widths 
do not coincide with BPs (see Fig.~\ref{widths}). 
The regions of largest \obsline\ line width 
coincide with the largest blueshifts. 
One may naively think that 
these regions present
systematic upflows of
a few km~s$^{-1}$, but they really
contain 
much larger
spatially unresolved upflows and downflows.
The line widths are much larger than the line shifts, and
a typical width of
0.02~nm corresponds to 60 km~s$^{-1}$
at \obsline. 
%
%
%

\subsection{Explosive events}\label{results3}
\begin{figure*}
\centering
\includegraphics[angle=90.,width=0.9\textwidth]{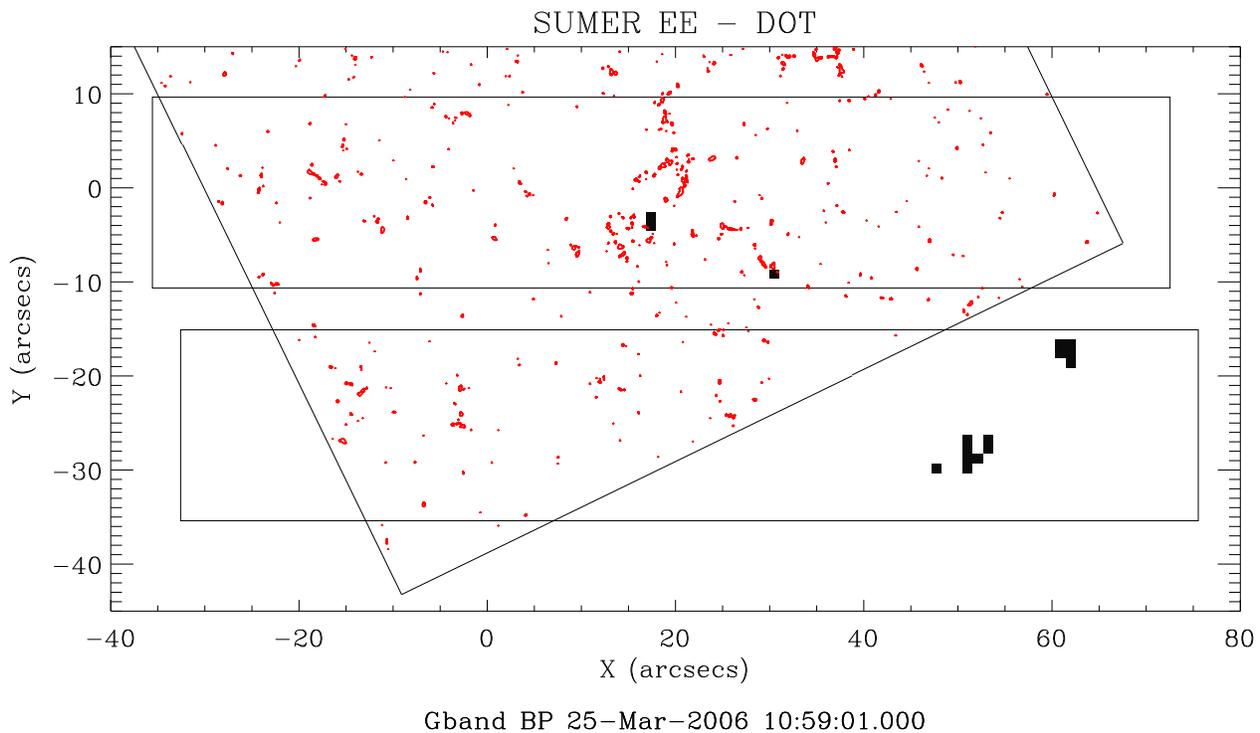}
\caption{
Image of the pixels with EEs (non-Gaussian line profiles)
together with the G-band BPs. The EE pixels are represented in black. 
The rest of the layout remains
as for  Fig.~\ref{res3}.
}
\label{ee1}
\end{figure*}

Explosive Events (EEs) are believed to be the result of 
magnetic reconnection. 
They can be identified because the line profiles are 
strongly non-Gaussian, and they often coincide with locations
of very large Doppler shifts and/or line widths derived by 
single-Gaussian fitting
\citep[][]{der89,ter04}.
We run a procedure to identify EEs by comparing the result of a
single-Gaussian fit with the observed profiles. 
All profiles for which at least three contiguous spectral pixels 
consistently deviating by more than two sigma from the
fit are flagged as EE.  
In our case we find 26 such spectra 
clustered in at least 7 patches. Unfortunately, 
most EE are outside the DOT FOV -- 
the EEs selected in rasters 5 and 6 are 
represented  in Fig.~\ref{ee1}.
From this very reduced statistics, we find no 
clear
overlapping between EEs and G-band BPs,
although each EE is not very far from a
BP either.
The spatial separation between BPs and EEs 
does not seem to be 
an artifact due to the lack of simultaneity between the
SUMER spectra and the DOT image chosen to represent
the photosphere. (Both EEs and BPs are transitory
events lasting shorter than the SUMER scans. The 
DOT snapshot of best seeing may miss BPs existing
during the individual EEs, 
masking a putative relationship
between EEs and BPs.)
BPs  do not coincide with EEs even when
the G-band image closest in time to each EE is used to search for BPs.
Another possible bias has to do with the criterion to
select EEs, which may overlook some of the weaker events or events
that happen to produce a very broad but fairly Gaussian profile. However,
as pointed out in
\S~\ref{results2}, BPs also avoid regions of large Doppler widths or
shifts, which reinforces the spatial disconnection
between BPs and EEs.  
%
%

\section{Discussion}\label{conclusions}
As far as we are aware of, this work represents the first
attempt to find footpoints of quiet Sun TR loops 
in the interior of supergranulation cells. G-band 
bright points (BPs) are used as proxies for 
photospheric magnetic field concentrations which
may anchor and guide the TR \obsline\ structures 
simultaneously observed with SUMER. As we anticipated, 
most photospheric magnetic concentrations do not appear in 
MDI magnetograms, therefore, traditional measurements 
miss TR loop footpoint candidates. Figure~\ref{res3}
presents our best effort to provide 
SUMER \obsline\ radiances co-aligned with the 
underlying photospheric magnetic structures. Unfortunately,
these and other similar maps do not allow us to 
uniquely identify G-band BPs with particular
TR loops.
Part of the difficulties have to do with identifying 
intrinsically fuzzy forms
in our reduced FOV
(\S~\ref{sumerobs}).
However, loosely using the term {\em loop} to denote the 
elongated structures present in the \obsline\ maps,
we find that some of them 
have BPs in the extremes,
but some other BPs are not obviously  associated with
specific TR structures. What we find, however, is a series of 
new observational properties linking the BPs with the
TR structures. 
In this sense the paper represents a study of the
relationships between the TR observed in \obsline\ 
and the presence of BPs.
The observed properties are not inconsistent
with the notion that BPs are indeed
footpoints of TR loops, but to go beyond this conjecture
is not free from speculation.

The observed G-band BPs are associated with 
bright \obsline\ structures, although there may be a 
tendency for the BPs to avoid the brightest central parts 
of the TR structures  (see Fig.~\ref{res3} and 
the dashed line in Fig.~\ref{histogram1}b).
In particular, the brightest structure in our FOV 
(Fig.~\ref{res3}, $X\simeq 15$\arcsec\ and $Y\simeq 0$\arcsec) 
is surrounded by two chains of BPs. It is tempting 
to think of the \obsline\ structure as system 
of loops joining the  two observed chains of BPs.
However, this interpretation is inconsistent with the fact that
the region containing the BPs and the \obsline\ structure
appears unipolar on the MDI magnetogram
(see Fig.~\ref{gband4}). 
A possible way to reconcile the existence of a system
of closed loops with the MDI magnetogram is the presence of spatially 
unresolved mixed polarities,
the MDI polarity being the dominating one.
(The existence of 
unresolved mixed polarities in the quiet Sun 
is both, expected from numerical simulations, \citealt{cat99a}, and 
observed, \citealt{san00}.) 
Alternatively, the BPs surrounding
the brightest \obsline\ structure may be footpoints 
of a bunch of magnetic field lines forming a large multi-strand loop which 
closes down outside the FOV. 
In this case, however, one would have to 
understand why the BPs avoid the brightest area.  
One possibility would be that magnetic 
structures are present, but they do not show up as BPs
because the field strength is too low, the intergranular 
lanes too cold, etc. \citep[see][]{san01}. Then the question
of why the magnetic structures at this particular
location have peculiar properties remains.

The TR loop model by \citet{spa06} satisfactorily
reproduces two non-trivial observables of the quiet Sun TR, namely,
the emission measure versus temperature distribution,
and the temperature dependence of the persistent redshifts. 
According to this model the
spectral lines forming at temperatures similar to 
that of \obsline\ ($\sim8\times 10^{4}$~K) are
systematically redshifted.
The corresponding plasma velocities turn out to be
in the range between 8 and 15 km~s$^{-1}$. 
We find that the position of the \obsline\ line profiles measured in the
G-band BPs exhibit redshifts
with respect to its rest wavelength corresponding to average plasma
velocities of 13$\pm 5$~km~s$^{-1}$ (Fig.~\ref{histogram1}),
in good agreement with the predictions of the
TR loop model. Moreover, \obsline\ spectra appear to be systematically
redshifted by about 2~km~s$^{-1}$
with respect to the average 
TR (\S~\ref{results2}).  These results are consistent
with the BPs being footpoints of loop-like structures.

Explosive Events (EEs) are believed to be the result of 
magnetic reconnection. Where the reconnection takes place is 
under debate. There are authors proposing for 
reconnection in either the photosphere/chromosphere
(the reconnection in the photosphere leads to shocks that accelerate the 
plasma at TR temperatures, \citeauthor{tar00} \citeyear{tar00}),
the TR (direct formation of bi-directional jets at the 
reconnection site, e.g., \citeauthor{inn97} \citeyear{inn97}), and the corona 
(reconnection high in the corona generates high energy particle 
beams that heat and accelerate the chromospheric plasma leading to the 
TR signature, \citeauthor{benz99} \citeyear{benz99}). 
We have identified several
of those events in our FOV, finding a tendency to avoid
BPs. This is a new result suggesting that EEs are not
located low in the chromosphere (at least not lower than the point 
where the field starts expanding significantly). 
If BPs are footpoints  of 
loops undergoing reconnection, then the fact that the EEs 
do not coincide with them indicates that site of TR plasma acceleration
is far from the photospheric footpoints. This would exclude the hypothesis of a
flare-like mechanism, as the particle beams reaching the loop footpoints would
result in plasma accelerated at, or very close to, the footpoints.
In case of shocks formed by reconnection in the photosphere, these shocks 
travel more than one Mm (distance between the observed EEs and the closest BPs)
before being dissipated.
Reconnection and consequent heating and plasma acceleration in the low 
chromosphere seems also excluded.
However, it should be mentioned that EEs have a weak
signature in chromospheric lines \citep{ter02} and that there is some evidence
that EEs are first  observed in chromospheric lines, and then in TR lines
\citep{mad02}.

As mentioned 
at the beginning of the section, this work represents
a first attempt to
outline research avenues for
identifying the 
footpoints of quiet Sun TR loops.
Clearly, the temporal 
and the angular resolution of both the visible and the UV data must 
be improved to 
proceed further.
Observations obtained with SUMER, DOT and instruments on the
satellite Hinode\footnote{\tt http://solar-b.nao.ac.jp/index\_e.shtml}
are expected to yield such improvements.

\acknowledgements
The authors acknowledge the use of the Solar Soft package for
data reduction and analysis.
This work has been 
partly funded by the Spanish Ministry of Education and Science
(AYA2004-05792), and by the Italian Space Agency (ASI I/035/05/0).
The SUMER instrument and its operation are financed by the Deutsches 
Zentrum f\"{u}r Luft- und Raumfahrt (DLR), the Centre National d'\'Etudes 
Spatiales (CNES), the National Aeronautics and Space Administration (NASA), 
and the European Space Agency's (ESA) PRODEX programme (Swiss contribution). 
The instrument is part of ESA's and NASA's Solar and Heliospheric Observatory 
(SoHO). 
The DOT is operated by Utrecht University at the Observatorio 
del Roque de los Muchachos of the IAC. 
%

\bibliographystyle{aa} 

%
%
\end{document}